\documentclass[pre,twocolumn,showpacs,preprintnumbers,amsmath,amssymb]{revtex4}
\usepackage{graphicx}
\usepackage{dcolumn}
\usepackage{bm}
\def \bea{\begin{eqnarray}}
\def \eea{\end{eqnarray}}
\def \non{\nonumber}
\begin{document}


\title{Isolated Non-Equilibrium Systems in Contact}
\author{Yair Shokef$^1$}
\thanks{Formerly: Srebro}
\author{Gal Shulkind$^2$}
\author{Dov Levine$^2$}
\affiliation{$^1$Department of Physics and Astronomy, University of Pennsylvania, Philadelphia, PA 19104, USA\\
$^2$Department of Physics, Technion, Haifa 32000, Israel}
\date{\today}

\begin{abstract}

We investigate a solvable model for energy conserving non-equilibrium steady states. The time-reversal asymmetry of the dynamics leads to the violation of detailed balance and to ergodicity breaking, as manifested by the presence of dynamically inaccessible states. Two such systems in contact do not reach the same effective temperature if standard definitions are used. However, we identify the effective temperature that controls energy flow. Although this operational temperature does reach a common value upon contact, the total entropy of the joint system can decrease.

\end{abstract}


\maketitle


{\it What happens when systems in non-equilibrium steady states are brought into contact?} For contact between equilibrium systems, applying the second law of thermodynamics requires them to be {\it closed} or {\it isolated}, with a fixed total energy. When energy is allowed to flow between systems that are otherwise isolated from the environment, entropy maximization determines the final equilibrium state and leads to temperature equalization. Here, we study what happens when systems in far from equilibrium steady states are connected and allowed to reach a mutual non-equilibrium steady state.

Typically, non-equilibrium steady states are realized in driven systems, receiving a perpetual supply of energy which is removed by dissipation or contact to a thermal bath. These are clearly not isolated, making interpreting questions about contact between them more difficult. Nonetheless, one may question to what extent may non-equilibrium steady states be described by effective temperatures determining their properties \cite{teff,ITP}. 

The question of contact between far from equilibrium systems is motivated by striking differences between equilibrium systems and open, dissipative systems in contact, such as the breakdown of energy equipartition in driven granular gas mixtures \cite{gran_gas}. Additionally, the direction of energy flow and possible equalization of effective temperatures are of particular relevance in attaching a thermometer to a non-equilibrium system \cite{teff,baranyai_2000,garriga_ritort_2001b,berthier_barrat_2002,hatano_jou_2003}.

Soft matter systems comprised of macroscopic grains, colloids, foam bubbles, or living cells undergo irreversible dynamics. Contact between such driven dissipative systems differs from the equilibrium situation in two aspects: (i) the microscopic dynamics are irreversible, thus violating detailed balance (DB), and (ii) energy flows into the systems due to driving and out by dissipation \cite{footnote_open}. As a step toward obtaining a theoretical understanding of this complicated situation, we focus on the contact between isolated non-equilibrium systems undergoing microscopically irreversible but energy conserving dynamics \cite{footnote_particles}. We furthermore expect that this concept of an isolated non-equilibrium system may prove useful when studying a slowly degrading open system on short time scales.

In the first part of this Rapid Communication, we introduce and study a stochastic model for isolated systems which achieves a non-equilibrium steady state. The model exhibits a dynamics-induced hole of inaccessible states. This results from DB violation, which we quantify by means of the total probability current. The system is furthermore characterized by moments of the energy distribution, which we relate to various definitions of effective temperatures.

In the second part of this Rapid Communication, we consider contact between two such systems $A$ and $B$. For weak coupling between them, they affect each other only by partitioning energy between them, with the total energy $E_A + E_B$ being strictly fixed. In this limit, our model satisfies the zeroth law of thermodynamics: if two systems are in mutual steady states with a third system, they will also be in a mutual steady state one with the other. This is not the case for open driven dissipative systems.

Effective temperatures generally differ between the connected systems (as in \cite{ITP,gran_gas}). We are able, however, to identify an effective {\it operational} temperature \cite{baranyai_2000,hatano_jou_2003} that determines energy flow and equalizes at contact. This operational temperature is a property of an individual system, and does not depend on the coupling details between the systems. This notwithstanding, in contrast to the second law of thermodynamics, entropy defined in the standard way is not maximized upon contact.
  

Our model consists of $N$ particles with positive energies $\{ e_i \}_{1 \leq i \leq N}$, and is a modification of the open model of \cite{srebro_levine_2004,shokef_levine_2006}. At each time step three particles $i \ne j \ne k$ in the system are chosen at random. The first two ($i$, $j$) exchange energy ``dissipatively'' with a constant restitution coefficient $0 \leq \alpha \leq 1$, and the third ($k$) receives the energy dissipated from them. The non-dissipated energy $\alpha (e_i+e_j)$ is randomly redistributed between the first two. Thus, the particle energies after the interaction are:
\bea
e'_i &=& z \alpha (e_i+e_j) \non \\
e'_j &=& (1-z) \alpha (e_i+e_j) \non \\
e'_k &=& e_k + (1-\alpha) (e_i+e_j) \label{eq:model} ,
\eea
where $0 \leq z \leq 1$ is drawn randomly at every interaction from a uniform distribution, and the energies of all other particles in the system remain unchanged. We note that when $\alpha = 1$, the system is in equilibrium. For $\alpha\!\!<\!\!1$, although the total energy in an isolated system $E = \Sigma_{i=1}^{N}e_i$ remains exactly constant, its steady state is far from equilibrium because of the time-reversal asymmetry of the dynamics. We shall demonstrate this through the system's non-ergodicity and DB violation \cite{footnote_bdd}. 

\begin{figure}[t]
\includegraphics[width=6.8cm]{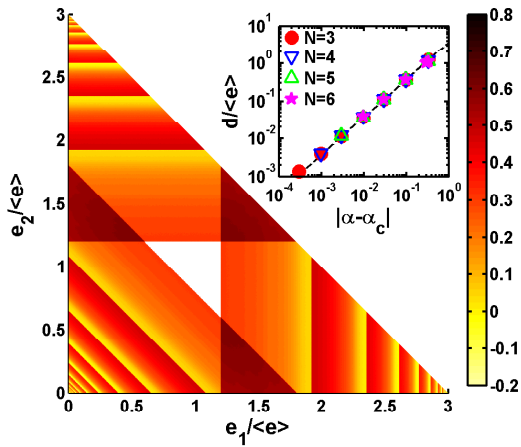}
\caption{\label{fig:slice_N3_a0.6} (Color online) Configuration space occupation for $N=3$, $\alpha=0.6$. Colorbar indicates the log of the occupation probability density. The triangular hole in the middle corresponds to inaccessible states. Inset: Distance to equal energy state. Dashed line is the theoretical prediction of Eq. (\ref{eq:dist}).}
\end{figure}

For $\alpha < \frac{2}{3}$ not all energetically allowed states are dynamically accessible. If the system starts in an inaccessible state, the dynamics will never return it there. Nonetheless, the system has a unique steady state and all accessible states are dynamically connected: the system can evolve with time between any two of them, and ensemble averages coincide with time averages. For simplicity of visualization, we first consider a system of $N=3$ particles. For $\alpha=1$ our model is equivalent to an equilibrium system with a flat single particle density of states (DOS) \cite{srebro_levine_2005}, and we observe a uniform occupation of the constant energy surface in configuration space. For $\alpha<1$ we find numerically a non-uniform occupation of configuration space. This alone does not imply non-equilibrium behavior in the absence of a DOS (or phase space measure) for non-Hamiltonian dynamics \cite{srebro_levine_2005}. For $\alpha$ beneath the critical value $\alpha_{c} = \frac{2}{3}$, a region of dynamically inaccessible states appears (see Fig. \ref{fig:slice_N3_a0.6}), which may not be explained by a DOS. This is seen if the system is coupled to a reservoir, when these states become accessible and populated. As $\alpha$ decreases, the inaccessible states take over more of configuration space, until at the maximally dissipative limit ($\alpha=0$) the only occupied states are with a single particle holding all the system's energy.  

We now calculate the threshold in $\alpha$ for inaccessible states to exist. As $\alpha$ decreases below $\alpha_{c}$, the first state to become inaccessible is that where all particles have equal energies. For smaller values of $\alpha$, the inaccessible region grows around this equal energy state. For $N=3$ we substitute the final energies $e'_1=e'_2=e'_3=\langle e \rangle \equiv \frac{E}{N}$ in Eq. (\ref{eq:model}), and see that the energy of particle $k$ prior to the last interaction had to be $e_k = \left( 3 - \frac{2}{\alpha} \right) \langle e \rangle$. All energies remain positive by the dynamics, hence in order for this $e_k$ to be positive, $\alpha$ must be larger than $\alpha_c = \frac{2}{3}$. We therefore conclude that when $\alpha < \alpha_c$ the state $e_1=e_2=e_3$ may not be reached by the dynamics, although it is energetically allowed. For $N>3$ we similarly check what is the initial state required for the system to reach $e'_1 = \ldots = e'_N = \langle e \rangle$ after a single interaction. Again, Eq. (\ref{eq:model}) implies that the energy of particle $k$ before the interaction is $e_k = \left( 3 - \frac{2}{\alpha} \right) \langle e \rangle$, leading to the same relation as above. To conclude, due to the irreversibility of the dynamics an isolated system has inaccessible states for $\alpha<\alpha_{c}=\frac{2}{3}$, irrespective of the number of particles in the system.

To compute the size of the inaccessible hole in configuration space, we note that for $\alpha \lesssim \alpha_c$ the {\it accessible} state {\it closest} to the inaccessible equal energy state is that obtained from the pre-collisional state with $e_k=0$, $e_i+e_j=3 \langle e \rangle$, and all other particles having energy $\langle e \rangle$. The interaction then brings the system to $e_i=e_j=\frac{3 \alpha}{2} \langle e \rangle$, $e_k=3(1-\alpha) \langle e \rangle$. This is the nearest the system can get to the equal energy state, as measured by the distance:
\bea
d \equiv \left[ \sum_{i=1}^N \left( e_i - \langle e \rangle \right)^2 \right]^{1/2} = \sqrt{\frac{3}{2}} \left( \frac{2}{3} - \alpha \right) \langle e \rangle. \label{eq:dist}
\eea
The inset to Fig. \ref{fig:slice_N3_a0.6} verifies this form for the minimal distance to the equal energy state for general $N$.


\begin{figure}[b]
\includegraphics[width=5.1cm]{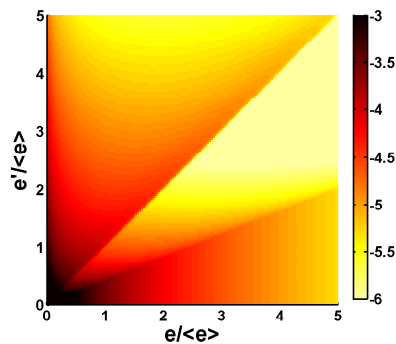}
\caption{\label{fig:db} (Color online) Transition rates $R(e,e')$ of a single particle from energy $e$ to energy $e'$ for $\alpha=0.4$. Colorbar indicates the log of the transition rate.}
\end{figure}

\begin{figure}[t]
\includegraphics[width=8.3cm]{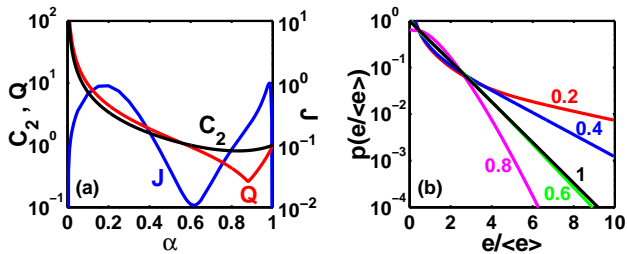}
\caption{\label{fig:db_vs_a} (Color online) a) Total probability current $J$, tail decay constant $Q$, and normalized second moment $C_2$ vs $\alpha$. b) Single particle energy distribution for various values of the restitution coefficient $\alpha$ (indicated next to each line).}
\end{figure}

Our dynamics are irreversible, and it is of interest to quantify the resulting DB violation, which may serve as a measure of how far a system is from equilibrium. To this end, we consider a single particle in a large isolated system. Figure \ref{fig:db} depicts the particle's steady state transition rates $R(e,e')dede'$ from a region $de$ around $e$ to a region $de'$ around $e'$ \cite{footnote_rates}. In equilibrium these are symmetric irrespective of the DOS, that is $R(e,e')=R(e',e)$ for any pair of energies. In our case, $R(e,e') \ne R(e',e)$, and we quantify the overall DB violation by the total probability current defined as $J \equiv \int \left[ R(e,e') - R(e',e) \right]^2 de de'$ \cite{zia_schmittmann_2006}. $J$ vanishes in the equilibrium limit $\alpha=1$ and in the singular limit $\alpha=0$ (where transition rates between the extreme states of a single particle possessing all the energy are symmetric). It exhibits two peaks with a minimal DB violation around $\alpha \approx 0.6$ (see Fig. \ref{fig:db_vs_a}a) \cite{footnote_J_noise}.

Interestingly, the single particle energy distribution $p(e)$ exhibits minimal deviation from the Boltzmann distribution around $\alpha \approx 0.6$ as well. We use the procedure described in \cite{srebro_levine_2004,shokef_levine_2006} to calculate $p(e)$ in terms of all its moments $\langle e^n \rangle$, and find that each normalized moment $C_n \equiv \frac{\langle e^n \rangle}{n!\langle e \rangle^n}$ may be expressed in terms of lower moments:
\bea
C_n = \frac{\sum_{m=1}^{n-1} C_{n-m} \left[ K_n C_m + (1-\alpha)^m \sum_{\ell=0}^m C_{m-\ell}C_{\ell} \right]}{2(1-K_n)}
, \label{eq:mom_recur}
\eea
with $K_n \equiv \frac{2\alpha^n}{n+1} + (1-\alpha)^n$. The distributions for different values of $\alpha$ have exponential high energy tails with different decay constants (see Fig. \ref{fig:db_vs_a}b), but share the same first moment $\langle e \rangle$. As in \cite{shokef_levine_2006}, we characterize the distribution's high energy tail by ratios of succeeding high order moments. For a $p(e) \sim \exp[-e/(Q \langle e \rangle)]$ tail, as $n \rightarrow \infty$, one has $C_n/C_{n-1} \rightarrow Q$ with $Q$ finite. Iterating Eq. (\ref{eq:mom_recur}) we find that $C_n/C_{n-1}$ reaches a constant value $Q(\alpha)$ which switches from $>1$ to $<1$ at $\alpha \approx 0.58$ (see Fig. \ref{fig:db_vs_a}a). 

A measure for deviation from exponential behavior at moderate energies is the normalized second moment
\bea
C_2 = \frac{\langle e^2 \rangle}{2\langle e \rangle^2} = \frac{9 - 12 \alpha + 5 \alpha^2}{12 \alpha - 10 \alpha^2} . \label{eq:C2}
\eea
$C_2$ switches from $>1$ to $<1$ at $\alpha=0.6$ (see Fig. \ref{fig:db_vs_a}a). For our model it is equal to the ratio of two effective temperatures in the system $C_2 = T_{FD} / T_G$ \cite{srebro_levine_2004,shokef_levine_2006,shokef_bunin_levine_2006}: the fluctuation-dissipation temperature $T_{FD}$ scales fluctuations around the steady state to corresponding response functions; the granular temperature $T_G \equiv \langle e \rangle$ measures the average energy per degree of freedom.


We now consider thermal contact between two systems, $A$ and $B$, comprised of particles of different types. We assume the irreversible dynamics of Eq. (\ref{eq:model}) with restitution coefficients $\alpha_{AA}$, $\alpha_{BB}$, and $\alpha_{AB}=\alpha_{BA}$ depending on the species of the first two particles ($i$, $j$) in each interaction. For example, when an $A$ particle and a $B$ particle collide, the restitution coefficient is $\alpha_{AB}$ irrespective of the type of particle $k$ receiving the dissipated energy.

We define a dimensionless coupling strength $f$ between the systems, so that $f=0$ corresponds to no contact and $f=1$ to intimate contact as in a homogeneous mixture. We consider the case where the two systems have the same number of particles $N$ and exhibit the same interaction rate per particle. Therefore, of the eight possible ways to choose three particles from the two systems, each of the six inter-system interactions ($AAB$, $ABA$, $ABB$, $BAA$, $BAB$, $BBA$) occur with probability $f/8$, and each of the two intra-system interactions ($AAA$, $BBB$) with probability $(4-3f)/8$. We solve the model for general $f$ and concentrate here on the weak coupling limit $0 < f \ll 1$.


Averaging the energy transfers between the systems over these possible interactions we find that 
\bea
\frac{d \langle e_A \rangle}{dt} = \frac{f}{4} \left[ \left( 2 - \alpha_{BB} \right) \langle e_B \rangle - \left( 2 - \alpha_{AA} \right) \langle e_A \rangle \right] \label{eq:dedt} .
\eea
This enables us to identify $T_O \equiv (2 - \alpha) \langle e \rangle$ as the operational temperature \cite{baranyai_2000,hatano_jou_2003} of our model: heat flows from the system of high $T_O$ to the one with low $T_O$ until they eventually equalize. Due to the energetic isolation of our model, $T_O$ is defined for the individual systems and, unlike open systems \cite{hatano_jou_2003}, depends only on the system's properties and not on the contact details ($f$ and $\alpha_{AB}$). 

$T_O$ of a non-equilibrium system may be measured by attaching to it a thermometer with $\alpha=1$. A small enough thermometer will not affect the system. After the systems reach a mutual steady state we detach the thermometer from the system and allow it to reach equilibrium. Its (true) temperature then is equal to the operational temperature of the measured system since in the equilibrium limit $\alpha=1$, $T_O$ gives the temperature.

Other definitions of effective temperatures do not equalize upon contact and hence do not behave as thermodynamic temperatures: The granular temperature satisfies $T_G \equiv \langle e \rangle = T_O / (2 - \alpha)$, and thus differs across the systems; The fluctuation-dissipation temperatures in each of the two systems in contact may be calculated for general coupling, and shown to depend on the coupling details. In the weak coupling limit the contact affects only the partitioning of energy between the systems and not the normalized energy distributions within each of them. Therefore, in each system $T_{FD}$ may be evaluated from $T_G$ by Eq. (\ref{eq:C2}) and using $C_2 = T_{FD} / T_G$. This yields different values of $T_{FD}$ in the two systems.


In equilibrium, temperature equalization reflects the more fundamental entropy maximization. The equalization of an effective temperature $T_O$ does not necessarily imply that entropy is maximized here. Therefore, we now test the second law of thermodynamics and consider what happens to the total entropy of two systems when they are connected. We begin with disconnected systems of different types of particles ($\alpha_{AA} \ne \alpha_{BB}$), each in its own isolated steady state, with some partitioning of energy between them. We then connect the two and allow them to reach a mutual steady state. At this time, we disconnect them and wait for each one to reach its new steady state. Long after they are disconnected, the effect of the contact is to have repartitioned the energy without modifying the normalized energy distributions in each of the systems, and each system's total energy is its only energy scale. We may therefore write the multi-particle energy distributions in the systems as \cite{shokef_levine_2006}
\bea
p_A(e_1,\ldots,e_N) &=& \left(E_A\right)^{-N} \varphi \left( \frac{e_1}{E_A},\ldots,\frac{e_N}{E_A};\alpha_{AA} \right) 
\non \\  
p_B(e_1,\ldots,e_N) &=& \left(E_B\right)^{-N} \varphi \left( \frac{e_1}{E_B},\ldots,\frac{e_N}{E_B};\alpha_{BB} \right) , 
\eea
with $\varphi$ a dimensionless function of the dimensionless energies $\{ e_i/E \}$ and the dimensionless model parameters ($\alpha_{AA}$ and $\alpha_{BB}$ in this case). The systems' entropies $S \!\! \equiv  \!\! - \int p(e_1,\ldots,e_N) \ln {p(e_1,\ldots,e_N)} de_1 \cdots de_N$ now scale as $S_A = N \ln (E_A) + const$, $S_B = N \ln (E_B) + const$, with the additive constants depending only on the normalized energy distribution within each system and not on the way energy is partitioned between the systems, that is, only on $\alpha_{AA}$ and $\alpha_{BB}$ but not on $E_A$ and $E_B$ \cite{footnote_TS}.

When the systems are disconnected entropy is unquestionably additive. Therefore, these expressions for the entropy imply that equipartition of energy between the systems would maximize the total entropy. Since, however, the irreversibility of the dynamics leads to $E_A \!\! \ne \!\! E_B$, we conclude that entropy is not maximized upon contact  \cite{footnote_dickman}. In particular, if two systems of equal energy per degree of freedom are connected, their entropy will decrease. That is, unlike the equilibrium case, the second law of thermodynamics does not hold and entropy defined in the standard way may decrease rather than increase when a constraint is removed \cite{footnote_entropy_decrease}. 


In conclusion, we characterized non-equilibrium steady states of an energy conserving model, and investigated what happens when two such systems are brought into thermal contact. Although effective temperatures generally do not equalize, we identified an operational temperature which determines heat flow, ultimately reaching a common value when mutual steady state is achieved. Moreover, we demonstrated explicitly that the total entropy may decrease in this case.


We thank Robert Dorfman, Yariv Kafri, Andrea Liu, Alexander Lobkovsky and Fred MacKintosh for helpful discussions. This work was supported by Grant No. 660/05 of the Israel Science Foundation, the Fund for the Promotion of Research at the Technion, and NSF MRSEC program under DMR 05-20020.




\begin{thebibliography}{10}
\bibitem{footnote_rates}
We scale time so that $\int R(e,e') de de' = 1$; The transition rate $R(e,e')$ from $e$ to $e'$ should not be confused with the conditional probability per unit time $W(e,e')$ for a transition to $e'$ given that the system is at $e$ [in steady state these are related by $R(e,e')=W(e,e')p(e)$].
\bibitem{footnote_J_noise}
The system is not effectively in equilibrium at $\alpha \approx 0.6$ as the $10^{-2}$ value of $J$ there is significantly larger than the numerical noise of $4 \cdot 10^{-8}$ observed at $\alpha=1$.
\bibitem{footnote_entropy_decrease}
Entropy may decrease in any system with an inhomogeneous limiting distribution if started with a more homogeneous state. Our result is stronger since we start at a steady state and entropy decreases because the steady state reached after removing a constraint is less homogeneous than the initial one.
\end{thebibliography}
\end{document}